\documentclass[hidelinks,10pt,twocolumn]{article}
\usepackage{simpleConference}
\usepackage[pdftex,
            pdfauthor={Martin Czygan, Helge Holzmann, Bryan Newbold},
            pdftitle={Refcat: The Internet Archive Scholar Citation Graph},
            pdfsubject={Citation Graph},
            pdfkeywords={Citation Graph, Scholarly Communications, Web Archiving},
            pdfproducer={Latex with hyperref},
            pdfcreator={pdflatex}]{hyperref}
\usepackage[utf8]{inputenc}
\usepackage{times}
\usepackage{graphicx}
\usepackage{natbib}
\usepackage{doi}
\usepackage{amssymb}
\usepackage{url}
\usepackage{booktabs}       % professional-quality tables
\usepackage{amsfonts}       % blackboard math symbols
\usepackage{nicefrac}       % compact symbols for 1/2, etc.
\usepackage{caption}
\usepackage{datetime}
\usepackage{float}

\providecommand{\keywords}[1]{\textbf{\textit{Index terms---}} #1}
\begin{document}

\title{Refcat: The Internet Archive Scholar Citation Graph}

\author{Martin Czygan \\
	\\
	Internet Archive \\
	San Francisco, CA, USA \\
	martin@archive.org  \\
	\and
	Helge Holzmann \\
	\\
	Internet Archive \\
	San Francisco, CA, USA \\
	helge@archive.org  \\
	\and
	Bryan Newbold \\
	\\
	Internet Archive \\
	San Francisco, CA, USA \\
	bnewbold@archive.org  \\
	\\
}

\maketitle
\thispagestyle{empty}

\begin{abstract}
	As part of its scholarly data efforts, the Internet Archive (IA) releases a
	first version of a citation graph dataset, named \emph{refcat}, derived
	from scholarly publications and additional data sources. It is composed of
	data gathered by the fatcat cataloging
	project\footnote{\href{https://fatcat.wiki}{https://fatcat.wiki}} (the catalog that underpins IA Scholar), related
	web-scale crawls targeting primary and secondary scholarly outputs, as well
	as metadata from the Open
	Library\footnote{\href{https://openlibrary.org}{https://openlibrary.org}}
	project and
	Wikipedia\footnote{\href{https://wikipedia.org}{https://wikipedia.org}}.
	This first version of the graph consists of over 1.3B citations. We release
	this dataset under a CC0 Public Domain Dedication, accessible through
	Internet
	Archive\footnote{\href{https://archive.org/details/refcat\_2021-07-28}{https://archive.org/details/refcat\_2021-07-28}}.
	The source code used for the derivation process, including exact and fuzzy
	citation matching, is released under an MIT
	license\footnote{\href{https://gitlab.com/internetarchive/refcat}{https://gitlab.com/internetarchive/refcat}}.
	The goal of this report is to describe briefly the current contents and the
	derivation of the dataset.
\end{abstract}

\keywords{Citation Graph, Web Archiving}

\section{Introduction}

The Internet Archive released a first version of a citation graph dataset
derived from a corpus of about 2.5B raw references\footnote{Number of raw
	references: 2,507,793,772} gathered from 63,296,308 metadata records (which are
collected from various sources or based on data obtained by PDF extraction and
annotation tools such as GROBID~\citep{lopez2009grobid}). Additionally, we
consider integration with metadata from Open Library and Wikipedia.  We expect
this dataset to be iterated upon, with changes both in content and processing.

According to~\citep{jinha_2010} over 50M scholarly articles have been published
(from 1726) up to 2009, with the rate of publications on the
rise~\citep{landhuis_2016}. In 2014, a study based on academic search engines
estimated that at least 114M English-language scholarly documents are
accessible on the web~\citep{khabsa_giles_2014}.

Modern citation indexes can be traced back to the early computing age, when
projects like the Science Citation Index (1955)~\citep{garfield2007evolution}
were first devised, living on in commercial knowledge bases today.  Open
alternatives were started such as the Open Citations Corpus (OCC) in 2010 - the
first version of which contained 6,325,178 individual
references~\citep{shotton2013publishing}. Other notable projects include
CiteSeer~\citep{giles1998citeseer}, CiteSeerX~\citep{wu2019citeseerx} and
CitEc\footnote{\href{https://citec.repec.org}{https://citec.repec.org}}. The
last decade has seen the emergence of more openly available, large scale
citation projects like Microsoft Academic~\citep{sinha2015overview} and the
Initiative for Open
Citations\footnote{\href{https://i4oc.org}{https://i4oc.org}}~\citep{shotton2018funders}.
In 2021, over one billion citations are publicly available, marking a ``tipping
point'' for this category of data~\citep{hutchins2021tipping}.

While a paper will often cite other papers, more citable entities exist such as
books or web links and within links a variety of targets, such as web pages,
reference entries, protocols or datasets. References can be extracted manually
or through more automated methods, by accessing relevant metadata or structured
data extraction from full text documents. Automated methods offer the benefits
of scalability. The completeness of bibliographic metadata in references ranges
from documents with one or more persistent identifiers to raw, potentially
unclean strings partially describing a scholarly artifact.

\section{Related Work}

Two typical problems in citation graph development are related to data
acquisition and citation matching. Data acquisition itself can take different
forms: bibliographic metadata can contain explicit reference data as provided
by publishers and aggregators; this data can be relatively consistent when
looked at per source, but may vary in style and comprehensiveness when looked
at as a whole. Another way of acquiring bibliographic metadata is to analyze a
source document, such as a PDF (or its text), directly. Tools in this category
are often based on conditional random fields~\citep{lafferty2001conditional}
and have been implemented in projects such as
ParsCit~\citep{councill2008parscit}, Cermine~\citep{tkaczyk2014cermine},
EXCITE~\citep{hosseini2019excite} or GROBID~\citep{lopez2009grobid}.

The problem of citation matching is relatively simple when common, persistent
identifiers are present in the data. Complications mount, when there is
\emph{Identity Uncertainty}, that is ``objects are not labeled with unique
identifiers or when those identifiers may not be perceived
perfectly''~\citep{pasula2003identity}. CiteSeer has been an early project
concerned with citation matching~\citep{giles1998citeseer}. A taxonomy of
potential issues common in the matching process has been compiled
by~\citep{olensky2016evaluation}.  Additional care is required, when the
citation matching process is done at scale~\citep{fedoryszak2013large}. The
problem of heterogenity has been discussed in the context of datasets
by~\citep{mathiak2015challenges}.

Projects and datasets centered around citations or containing citation data as
a core component are COCI, the ``OpenCitations Index of Crossref open
DOI-to-DOI citations'', which was first released
2018-07-29\footnote{\href{https://opencitations.net/download}{https://opencitations.net/download}}
and has been regularly updated~\citep{peroni2020opencitations}. The
WikiCite\footnote{\href{https://meta.wikimedia.org/wiki/WikiCite}{https://meta.wikimedia.org/wiki/WikiCite}}
project, ``a Wikimedia initiative to develop open citations and linked
bibliographic data to serve free knowledge'' continuously adds citations to its
database\footnote{\href{http://wikicite.org/statistics.html}{http://wikicite.org/statistics.html}}.
Microsoft Academic Graph~\citep{sinha2015overview} is comprised of a number of
entities\footnote{\href{https://docs.microsoft.com/en-us/academic-services/graph/reference-data-schema}{https://docs.microsoft.com/en-us/academic-services/graph/reference-data-schema}}
with \emph{PaperReferences} being one relation among many others.

% There are a few large scale citation dataset available today. COCI, the
% ``OpenCitations Index of Crossref open DOI-to-DOI citations'' was first
% released 2018-07-29. As of its most recent release\footnote{\url{https://opencitations.net/download}}, on
% 2021-07-29, it contains
% 1,094,394,688 citations across 65,835,422 bibliographic
% resources~\citep{peroni2020opencitations}.
%
% The WikiCite\footnote{\url{https://meta.wikimedia.org/wiki/WikiCite}} project,
% ``a Wikimedia initiative to develop open citations and linked bibliographic
% data to serve free knowledge'' continously adds citations to its database and
% as of 2021-06-28 tracks 253,719,394 citations across 39,994,937
% publications\footnote{\url{http://wikicite.org/statistics.html}}.
%
% Microsoft Academic Graph~\citep{sinha2015overview} is comprised of a number of
% entities\footnote{\url{https://docs.microsoft.com/en-us/academic-services/graph/reference-data-schema}}
% with \emph{PaperReferences} being one relation among many others. As of 2021-06-07\footnote{A recent copy has been preserved at
% 	\url{https://archive.org/details/mag-2021-06-07}}  the
% \emph{PaperReferences} relation contains 1,832,226,781 rows (edges) across 123,923,466
% bibliographic entities.
%
% Numerous other projects have been or are concerned with various aspects of
% citation discovery and curation as part their feature set, among them Semantic
% Scholar~\citep{fricke2018semantic}, CiteSeerX~\citep{li2006citeseerx} or Aminer~\citep{tang2016aminer}.
%
% As mentioned in~\citep{hutchins2021tipping}, the number of openly available
% citations is not expected to shrink in the future.

\section{Dataset}

We released the first version of the \emph{refcat} dataset in a format used
internally for storage and to serve queries (and which we call \emph{biblioref}
or \emph{bref} for short). The dataset includes metadata from fatcat (the
catalog underpinning IA Scholar), the Open Library project and inbound links
from the English Wikipedia.  The dataset is integrated into the
\href{https://fatcat.wiki}{fatcat.wiki website} and allows users to explore
inbound and outbound
references\footnote{\href{https://guide.fatcat.wiki/reference\_graph.html}{https://guide.fatcat.wiki/reference\_graph.html}}.

The format records source and target identifiers, a
few metadata attributes (such as year or release stage, i.e. preprint, version of record, etc) as well as
information about the match status and provenance.

The dataset currently contains 1,323,423,672 citations across 76,327,662
entities (55,123,635 unique source and 60,244,206 unique target work
identifiers; for 1,303,424,212 - or 98.49\% of all citations - we do have a DOI
for both source and target).  The majority of matches - 1,250,523,321 - is
established through identifier based matching (DOI, PMIC, PMCID, ARXIV, ISBN).
72,900,351 citations are established through fuzzy matching techniques, where
references did not contain identifiers\footnote{This not necessary mean that
	the records in question do not have an identifier; however if an identifier
	existed, it was not part of the raw reference.}.
Citations from the Open Citations' COCI corpus\footnote{Reference dataset COCI
	v11, released 2021-09-04,
	\href{http://opencitations.net/index/coci}{http://opencitations.net/index/coci}}
and \emph{refcat} overlap to the most part, as can be seen in~Table~\ref{table:cocicmp}.

\begin{table}[]
	\begin{center}
		\begin{tabular}{lll}
			\toprule
			\bf{Set}              &                        & \bf{Count}    \\

			\midrule
			COCIv11 (C)           &                        & 1,186,958,897 \\ % zstdcat -T0 6741422v11.csv.zst | pv -l | wc -l
			\emph{refcat-doi} (R) &                        & 1,303,424,212 \\ % zstdcat -T0 /magna/refcat/2021-07-28/BrefDOITable/date-2021-07-28.tsv.zst | pv -l | LC_ALL=C sort -T /sandcrawler-db/tmp-refcat/ -S70% -k3,4 -u | zstd -c -T0 > uniq_34.tsv.zst # LC_ALL=C wc -l uniq_34_doi_lower_sorted.csv
			C $\cap$ R            & overlap                & 1,046,438,515 \\
			C $\setminus$ R       & COCIv11 only           & 140,520,382   \\ %  86,854,309    \\
			R $\setminus$ C       & \emph{refcat-doi} only & 256,985,697   \\ % xxx 295,884,246
		\end{tabular}
		\vspace*{2mm}
		\caption{Comparison between Open Citations COCI corpus (v11,
			2021-09-04) and \emph{refcat-doi}, a subset of \emph{refcat} where
			entities have a known DOI. At least 150,727,673 (58.7\%) of the 256,985,697 references in
			\emph{refcat-doi} only record links within a specific dataset provider;
			here GBIF with DOI prefix: 10.15468.}
		\label{table:cocicmp}
	\end{center}
\end{table}

% zstdcat -T0 /magna/refcat/2021-07-28/BrefDOITable/date-2021-07-28.tsv.zst | pv -l | LC_ALL=C sort -T /sandcrawler-db/tmp-refcat/ -S70% -k3,4 -u | zstd -c -T0 > uniq_34.tsv.zst
% zstdcat -T0 uniq_34.tsv.zst | pv -l | LC_ALL=C cut -f3,4 | zstd -c -T0 > uniq_34_doi.tsv.zst
% find . -name "*.csv" | parallel -j 16 "LC_ALL=C grep -v ^oci, {} | LC_ALL=C cut -d, -f2,3" | pv -l | zstd -c -T0 > ../6741422v10_doi_only.csv.zst

% v11
% time zstdcat -T0 /magna/data/opencitations/6741422v11.csv.zst | cut -d, -f2,3 | tr '[:upper:]' '[:lower:]' | LC_ALL=C sort -S50% -T /sandcrawler-db/tmp-refcat | pv -l > 6741422v11_doi_lower.csv

% TODO: some more numbers on the structure

% * doi-to-doi
% * only source doi
% * only target doi
% * paper-to-book (OL)
% * wikipedia-to-paper (WI)

\begin{table}[]
	\begin{center}
		\begin{tabular}{ll}
			\toprule
			\bf{Edge type}      & \bf{Count}    \\
			\midrule
			doi-doi             & 1,303,424,212 \\
			target-open-library & 20,307,064    \\
			source-wikipedia    & 1,386,941     \\
		\end{tabular}
		\vspace*{2mm}
		\caption{Counts of classic DOI to DOI references as well as outbound
			references matched against Open Library as well as inbound references
			from the English Wikipedia.}
		\label{table:structure}
	\end{center}
\end{table}

We started to include non-traditional citations into the graph, such as links
to books included in Open Library and links from the English
Wikipedia to scholarly works. For links between Open Library we employ both
identifier based and fuzzy matching; for Wikipedia references we used a published dataset~\citep{harshdeep_singh_2020_3940692} and we are contributing
to upstream projects related to wikipedia citation extraction, such as
\emph{wikiciteparser}\footnote{\href{https://github.com/dissemin/wikiciteparser}{https://github.com/dissemin/wikiciteparser}}
to generate updates from recent Wikipedia dumps\footnote{Wikipedia dumps are available on a monthly basis from \href{https://dumps.wikimedia.org/}{https://dumps.wikimedia.org/}.}. Table~\ref{table:structure} lists the
counts for these links. Additionally, we are examining web links appearing in
references: after an initial cleaning procedure we currently find 25,405,592
web links\footnote{The cleaning process is necessary because OCR artifacts and
	other metadata issues exist in the data. Unfortunately, even after cleaning not
	all links will be in the form as originally intended by the authors.} in the
reference corpus, of which 4,828,283 (19\%) have been preserved as of August 2021 with an HTTP 200
status code in the Wayback
Machine\footnote{\href{https://archive.org/web/}{https://archive.org/web/}} of
the Internet Archive. As an upper bound - if we additionally include all redirection (HTTP
3XX) and server error status codes (HTTP 5XX) - we find a total of 14,306,019 (56.3\%) links preserved.

We ran a live URL check\footnote{All links accessed on 2021-10-04 and 2021-10-05.} over a sample of 364415 links which appear in the reference
corpus \emph{and} have a HTTP 200 status code archival copy in the Wayback Machine. Of the 364415 links we find 305476 (83.8\%) responding with an HTTP
200 OK, whereas the rest of the links yield a variety of HTTP status codes,
like 404, 403, 500 and others - resulting in about 16\% of the links in the
reference corpus preserved at the Internet Archive being currently inaccessible
on the web\footnote{We used the
	\href{https://github.com/miku/clinker}{https://github.com/miku/clinker}
	command line link checking tool.} - making targeted web crawling and
preservation of scholarly references a key activity for maintaining
citation integrity.

% unpigz -c fatcat-refs-urllist-2021-06-17_lookup-20210714045637.tsv.gz| LC_ALL=C grep -F ')/' | grep -c -E "\W200\W"

\section{System Design}

\subsection{Constraints}

The constraints for the system design are informed by the volume and the
variety of the data. The capability to run the whole graph derivation on a
single machine\footnote{We used a shared virtual server with 24 cores and 48G
	of main memory. The most memory-intensive part of the processing currently are
	the buffers set aside for \emph{GNU sort}.} was a minor goal as well. In
total, the raw inputs amount to a few terabytes of textual content, mostly
newline delimited JSON. More importantly, while the number of data fields is
low, certain documents are very partial with hundreds of different combinations
of available field values found in the raw reference data. This is most likely
caused by aggregators passing on reference data coming from hundreds of
sources, each of which not necessarily agreeing on a common granularity for
citation data and from artifacts of machine learning based structured data
extraction tools.

Each combination of fields may require a slightly different processing path.
For example, references with an Arxiv identifier can be processed differently
from references with only a title.

\subsection{Data Sources}

Reference data comes from two main sources: explicit bibliographic metadata and
PDF extraction. The bibliographic metadata is taken from fatcat, which itself
harvests and imports web accessible sources such as Crossref, Pubmed, Arxiv,
Datacite, DOAJ, dblp and others into its catalog (as the source permits, data
is processed continuously or in batches). Reference data from PDF documents has
been extracted with GROBID\footnote{GROBID
	\href{https://github.com/kermitt2/grobid/releases/tag/0.5.5}{v0.5.5}}, with the
TEI-XML results being cached locally in a key-value store accessible with an S3
API\footnote{Currently,
	\href{https://github.com/chrislusf/seaweedfs}{https://github.com/chrislusf/seaweedfs}
	is used}. Archived PDF documents result from dedicated web-scale crawls of
scholarly domains conducted with
multiple open-source crawler technologies created by the Internet Archive
and a variety of seed lists targeting journal
homepages, repositories, dataset providers, aggregators, web archives and other
venues. A processing pipeline merges catalog data from the primary database and
cached data from the key-value store and generates the set of about 2.5B
references records, which currently serve as an input for the citation graph
derivation pipeline.

\subsection{Methodology}

Overall, a map-reduce style~\citep{dean2010mapreduce} approach is
followed\footnote{While the operations are similar, the processing is not
	distributed but runs on a single machine. For space efficiency, zstd~\citep{collet2018zstandard} is used to compress raw data and derivations.}, which allows
for some
uniformity in the processing. We extract \emph{(key, document)} tuples (as
TSV) from the raw JSON data and sort by key. We then group documents with the
same key and apply a function on each group in order to generate
our target schema or perform
additional operations such as deduplication or fusion of matched and unmatched references for indexing.

The key derivation can be exact (via an identifier like DOI, PMID, etc) or
based on a value normalization, like ``slugifying'' a title string. For identifier
based matches we can generate the target schema directly.  For fuzzy matching
candidates, we pass possible match pairs through a verification procedure,
which is implemented for \emph{release entity}\footnote{\href{https://guide.fatcat.wiki/entity\_release.html}{https://guide.fatcat.wiki/entity\_release.html}.} pairs. This procedure is a
domain dependent rule based verification, able to identify different versions
of a publication, preprint-published pairs and documents, which are
are similar by various metrics calculated over title and author fields. The fuzzy matching
approach is applied on all reference documents without any identifier (a title is
currently required).

We currently implement performance sensitive parts in the
Go programming language\footnote{\href{https://golang.org/}{https://golang.org/}}, with various processing stages (e.g.
conversion, map, reduce, ...) represented by separate command line tools. A
thin task orchestration layer using the luigi
framework\footnote{\href{https://github.com/spotify/luigi}{https://github.com/spotify/luigi}~\citep{bernhardsson2018rouhani},
	which has been used in various scientific pipeline
	application, like~\citep{schulz2016use},~\citep{erdmann2017design},~\citep{lampa2019scipipe},~\citep{czygan2014design}
	and others.} allows for experimentation in the pipeline and for single command
derivations, as data dependencies are encoded with the help of the
orchestrator. Within the tasks, we also utilize classic platform tools such as
GNU \emph{sort}~\citep{mcilroy1971research}.

During a last processing step, we fuse reference matches and unmatched items
into a single, indexable file. This step includes deduplication of different
matching methods (e.g. prefer exact matches over fuzzy matches). This file is
indexed into a search index and serves both matched and unmatched references
for the web application, allowing for further collection of feedback on match
quality and possible improvements.

With a few schema conversions, fuzzy matching has been be applied to Wikipedia
articles and Open Library (edition) records as well. The aspect of precision
and recall are represented by the two stages: we are generous in the match
candidate generation phase in order to improve recall, but we are strict during
verification, in order to control precision. Quality assurance for verification is
implemented through a growing list of test cases of real examples from the catalog and
their expected or desired match status\footnote{The list can be found under:
	\href{https://gitlab.com/internetarchive/refcat/-/blob/master/skate/testdata/verify.csv}{https://gitlab.com/internetarchive/refcat/-/blob/master/skate/testdata/verify.csv}.
	It is helpful to keep this test suite independent of any specific programming language.}.

\section{Limitations and Future Work}

As with other datasets in this field we expect this dataset to be iterated upon.

\begin{itemize}
	\item The fatcat catalog updates its metadata
	      continuously\footnote{A changelog can currently be followed here:
		      \href{https://fatcat.wiki/changelog}{https://fatcat.wiki/changelog}.} and web crawls are conducted
	      regularly. Current processing pipelines cover raw reference snapshot
	      creation and derivation of the graph structure, which allows to rerun the
	      processing pipeline based on updated data as it becomes available.

	\item Metadata extraction from PDFs depends on supervised machine learning
	      models, which in turn depend on available training datasets. With additional crawls and
	      metadata available we hope to improve models used for metadata
	      extraction, improving yield and reducing data extraction artifacts in
	      the process.

	\item As of this version, a number of raw reference
	      docs remain unmatched, which means that neither exact nor fuzzy matching
	      has detected a link to a known entity. Metadata might be missing. However, parts of the data
	      will contain a reference to a catalogued entity, but in a specific,
	      dense and harder to recover form.

	\item The reference dataset contains millions of URLs and their integration
	      into the graph has been implemented as a prototype. A full implementation
	      requires a few data cleanup and normalization steps.
\end{itemize}

\section{Acknowledgements}

This work is partially supported by grants from the \emph{Andrew W. Mellon
	Foundation}, especially ''Ensuring the Persistent Access of Open Access Journal
Literature: Phase II`` (1910-07256, Jefferson Bailey, Principal Investigator).

\appendix

\section*{Appendix: Reference Relations}

\begin{figure}[h]
	\centering
	\includegraphics[width=0.45\textwidth]{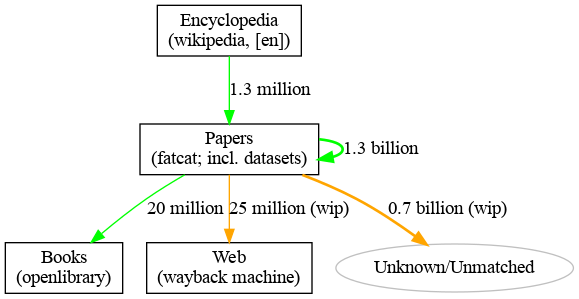}
	\caption{Schematics of the main reference entities; green: included in the
		corpus; orange: currently in development.}
	\label{fig:types}
\end{figure}

Figure~\ref{fig:types} shows the schematic reference relations.

% \pagebreak{}

\section*{Appendix: Data Quality}

A note on data quality: While we implement various data quality measures,
real-world data, especially coming from many different sources will contain
issues. Among other measures, we keep track of match reasons,
especially for fuzzy matching to be able to zoom in on systematic errors
more easily (see~Table~\ref{table:matches}).

\begin{table}[H]
	\footnotesize
	\captionsetup{font=normalsize}
	\begin{center}
		\begin{tabular}{@{}rlll@{}}
			\toprule
			\textbf{Count} & \textbf{Provenance} & \textbf{Status} & \textbf{Reason}      \\ \midrule
			934932865      & crossref            & exact           & doi                  \\
			151366108      & fatcat-datacite     & exact           & doi                  \\
			65345275       & fatcat-pubmed       & exact           & pmid                 \\
			48778607       & fuzzy               & strong          & jaccardauthors       \\
			42465250       & grobid              & exact           & doi                  \\
			29197902       & fatcat-pubmed       & exact           & doi                  \\
			19996327       & fatcat-crossref     & exact           & doi                  \\
			11996694       & fuzzy               & strong          & slugtitleauthormatch \\
			9157498        & fuzzy               & strong          & tokenizedauthors     \\
			3547594        & grobid              & exact           & arxiv                \\
			2310025        & fuzzy               & exact           & titleauthormatch     \\
			1496515        & grobid              & exact           & pmid                 \\
			680722         & crossref            & strong          & jaccardauthors       \\
			476331         & fuzzy               & strong          & versioneddoi         \\
			449271         & grobid              & exact           & isbn                 \\
			230645         & fatcat-crossref     & strong          & jaccardauthors       \\
			190578         & grobid              & strong          & jaccardauthors       \\
			156657         & crossref            & exact           & isbn                 \\
			123681         & fatcat-pubmed       & strong          & jaccardauthors       \\
			79328          & crossref            & exact           & arxiv                \\
			57414          & crossref            & strong          & tokenizedauthors     \\
			53480          & fuzzy               & strong          & pmiddoipair          \\
			52453          & fuzzy               & strong          & dataciterelatedid    \\
			47119          & grobid              & strong          & slugtitleauthormatch \\
			36774          & fuzzy               & strong          & arxivversion         \\
		\end{tabular}
		\vspace*{2mm}
		\caption{Table of match counts (top 25), reference provenance, match
			status and match reason. Provenance currently can name the raw
			origin (e.g. \emph{crossref}) or the method (e.g. \emph{fuzzy}). The match reason
			identifier encodes a specific rule in the domain dependent
			verification process and is included for completeness - we do not
			include the details of each rule in this report.}
		\label{table:matches}
	\end{center}
\end{table}

\bibliographystyle{plainnat}
\bibliography{main}
\end{document}